\documentclass[jgrga]{AGUTeX}

\usepackage[dvips]{graphicx}

\authorrunninghead{ZHANG et al}

\titlerunninghead{Analysis of space wave-particle coupling events}

\begin{document}

%
%

\title{Study of typical space wave-particle coupling events possibly related with seismic activity}
%

%
%


\authors{Zhenxia Zhang, \altaffilmark{1} ChenYu Wang,\altaffilmark{2} Xuhui Shen,\altaffilmark{3}
 Xinqiao Li,\altaffilmark{4} Shugui Wu,\altaffilmark{1}
 }

\altaffiltext{1}{National Earthquake Infrastructure Service, China
Earthquake Administration, Beijing, China. (zxzhang@neis.gov.cn).}
\altaffiltext{2}{Peking University, Beijing, China.}

\altaffiltext{3}{Institute of Earthquake Science,  China
Earthquake Administration, Beijing, China.}
\altaffiltext{4}{Institute of High Energy Physics, Chinese Academy
of Sciences, Beijing, China.}

%
%

\begin{abstract}
Based on the DEMETER satellite, we found two space wave-particle
coupling events during February 2010 taking place in the range of
McIlwain parameter $L$($1.27\sim1.37$). There are strong spatial
and temporal correlation between the particle bursts(PBs) and the
electromagnetic disturbances of the coupling events. The two PBs
show different energy spectrum characteristics, while the
corresponding electromagnetic disturbances concentrated on
different frequencies range. In agreement with the prediction of
the theory of wave-particle interaction, we conclude that the two
wave-particle interactions can be probably explained as following:
one is electron dominant precipitation with energy of
$0.09\sim0.2$ MeV induced by VLF electromagnetic wave with the
frequency of $14\sim20$ kHz, and another is proton dominant
precipitation with energy of $0.65\sim2.85$ MeV induced by VLF
electromagnetic wave with the frequency of $\le 100$ Hz. For the
first time, those particle bursts origin, from electron or proton
detected by the Instrument for the Detection of Particles(IDP) on
board, is inferred by theory calculation, although the Instrument
has no ability to identify the particle species.
\end{abstract}

\begin{article}

\section{Introduction}
There are many factors that caused the high energy particle
acceleration, precipitation and short-term, sharp particle
counting rates increase(particle bursts, or PBs), consisting of
ground-based VLF EM transmitter, lightning and thunderstorm,
ground nuclear test and blast, seismic activity, volcano natural
hazard and so on. The main mechanism is that the EM wave, radiated
from those events, couples with the energetic particle in the
radiation belt by wave-particle interaction. The interaction
accelerates the particle by changing the momentum of the particle,
or scatters the pitch angle of the particle and make them enter
the bounce or drift loss cone. So that a mass of energetic
particle accumulate at certain McIlwain parameter L value to
induce the particle precipitation, loss or particle burst.

Many effects of electron belts induced by man-made VLF
ground-based transmitters obey the theoretical model of
wave-particle interaction. Earlier, Inan {\it et al.} applied a
test-particle model of the gyroresonant wave-particle interaction
for calculating the precipitation characteristics of the particle
flux by a VLF transmitter(~\cite{Inan1985}). They found the
precipitation of the particle is controlled by the pitch angle
distribution near the edge of the loss cone. Horne {\it et
al.}(~\cite{Horne2005}) studied the mechanism of electron
acceleration by wave in the outer radiation belt. They show that
the electrons could be accelerated by electromagnetic waves at
frequencies of a few kHz, which can also increase the electron
flux by more than three orders of magnitude over the observed
timescale of one to two days. Based on DEMETER(Detection of
Electro-Magnetic Emissions Transmitted from Earthquake Regions)
satellite, Sauvaud and Maggiolo observed enhancements in the
~100-600keV drift-loss cone electron fluxes at L values between
$1.4\sim1.7$ induced by NWC transmitter(~\cite{sauvaud-NWC}). They
calculated the variation of the energy of enhanced electron fluxes
with L by first-order cyclotron resonance theory of wave-particle
interaction and obtained consistent results with the observation.
Graf {\it et al.} analyzed the energetic electron flux increase
near loss cone induced by ground-based VLF transmissions of NPM
observed via satellite-based detection, and compared the
precipitating flux with predictions based on ray-tracing analyses
of wave propagation and wave-particle
interaction(~\cite{Graf2009}). They indicated that the detection
rate is attributed to the orientation of the DEMETER particle
detector and obtained the agreement between the observations and
theory. Wang {\it et al.} explained the relationship between the
radiation belt electron precipitation observed by DEMETER and the
man-made VLF signals in the NPM experiment by the theory of
qusi-linear diffusion with resonant interaction, and indicated
that most of the non-correlation is caused by the restriction of
the pitch angle measurement of IDP (~\cite{wangping2011}). Li {\it
et al.} analyzed the energy spectrum of the NWC electron belts in
detail using on-off method and explained the pitch angle range
observed by quasi-linear diffusion equation of wave-particle
theory(~\cite{Lixinqiao2012}). The illustrations of precipitation
of inner radiation belt electrons, induced by VLF transmitters and
their explanation by wave-particle interaction, are also included
in the work of Inan {\it et al.}(~\cite{Inan2007}).

In addition to the man-made VLF transmitter, there are many
observations about the energetic electron precipitation and
particle bursts caused by seismic activity, observed by satellite
detectors. The representative works in this subject includes the
correlation between energetic particle burst and seismic activity
based on MARIA experiment studied by the
papers(~\cite{voronov1987,voronov1989,parrot2009}). Fidani and
Battistion performed a detailed analysis to the short-term and
sharp increases in high energy particle counting rates and their
correlations to seismic activity based on the whole period of ten
years burst activity from NOAA data (~\cite{Fidani2008}).

The DEMETER satellite with a ~700 km altitude, $98.3^{\circ}$
inclination orbit (~\cite{Parrot2006}) serves from June 2004 to
December 2012. An onboard instrument for particle
detection(IDP)(~\cite{Sauvaud2006}) measures 72.9keV$\sim2.35$MeV
electrons with 8.9 keV resolution in burst mode at one sample per
second, and the electric field instrument(ICE) measures electric
field fluctuations of $0\sim20$kHz in burst mode. Based on the
credible measurement of electric field and energetic particle
flux, we can study the particle precipitation and particle burst
events and the wave-particle interaction between them and the VLF
waves in detail.

When we look for the ionosphere disturbance in link of Chile
Earthquake on 27th February 2010, we found two particle burst
events and the corresponding electric field disturbance occurring
over epicenter based on Satellite
observation(~\cite{Zhangzhenxia2012}), with the figures shown in
Figure 1 and 2. The energy spectra of the PBs exhibits different
distribution structure for the two PBs.
In this paper, by calculating the quasi-linear diffusion
coefficients for field-aligned electromagnetic waves, we analyze
qualitatively the coupling properties of the two PBs and the
corresponding VLF electric spectrum disturbance according to the
information of their pitch angle, energy spectrum and frequency
range and so on.

\section{Observations}

The DEMETER satellite, with quasi-Sun-synchronous orbit, flies
downward (from north to south) during local daytime and flies
upward (from south to north) during local nighttime. The orbital
period is 102.86 minutes. Here we select the satellite data of
upward orbit to perform statistical analysis as the
electromagnetic wave can transmit into the ionosphere in night
more easier. We analyzed the high energy particle flux data sample
detected by IDP in the first three months of 2010. The charged
particle counting rates were considered in three energy regions,
90$\sim$600keV, 600$\sim$1000keV and 1000$\sim$2350keV,
respectively. We selected the satellites orbits which fly over the
epicenter region ($71^{\circ} 00^{\prime} W$, $30^{\circ}
00^{\prime} S$ ) within the longitude range of 10 degree and L
value of 0.1. So there are total 42 orbits across epicenter region
of Chile earthquake during seismic activity in the first three
months of 2010. Each orbit can fly across the epicenter region in
about several minutes.

\begin{figure}
\center{
 \noindent\includegraphics[width=18pc]{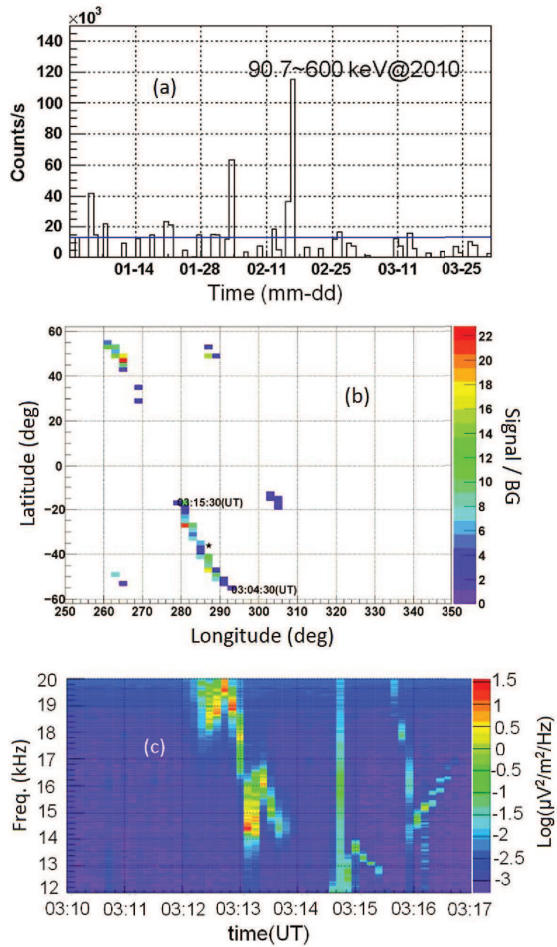}
 \caption{All observed figures for PB1 from orbit 30109 orbit on
16th February. (a) The distribution of
 averaged high energy charged particle counting rates in 2010.
 (b)The distribution of position and the duration UT of the
 energetic particle burst in which the value of Signal over
 average background is equal or larger than 2. The black star denotes the position of the
 epicenter of Chile Earthquake. (c)The disturbance of the VLF electric spectrum in ionosphere over
the epicenter detected by the DEMETER satellite from the same
orbit 30109. \label{f30109}
 }
}
 \end{figure}

\begin{figure*}
\center{
 \noindent\includegraphics[width=30pc]{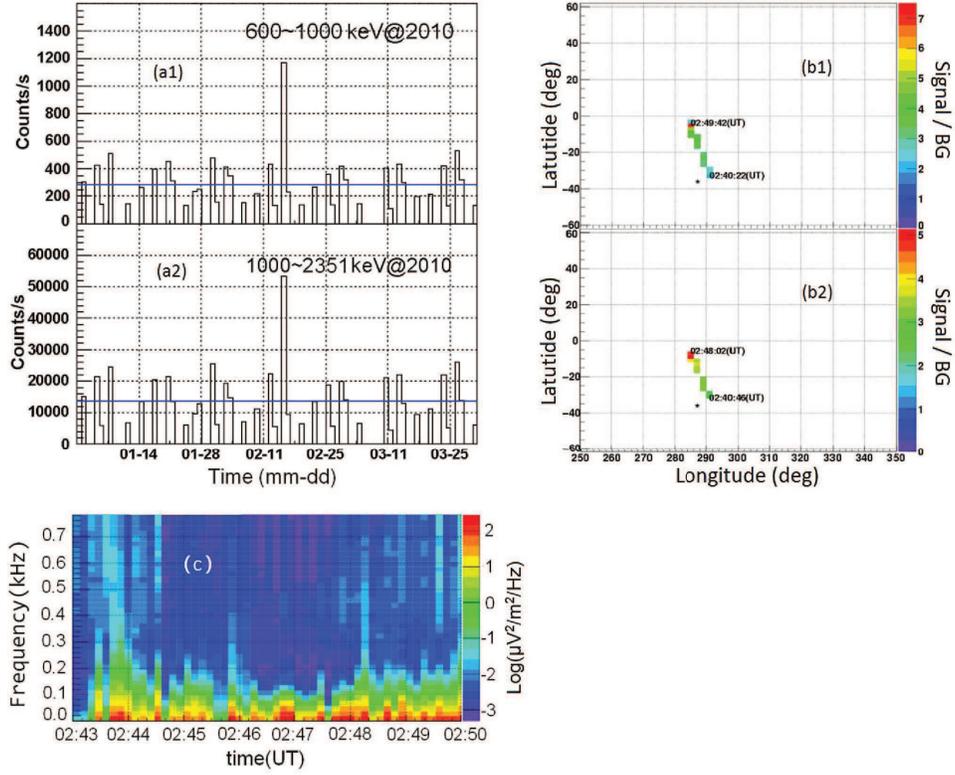}
 \caption{All observed figures for PB2 from orbit 30094 orbit on
15th February. Plots of (a1) and (a2) denotes the distribution of
 averaged high energy charged particle counting rates.
 Plots of (b1) and (b2) denotes the distribution of position and the duration UT of the
 energetic particle burst, for $600\sim1000keV$ and $1000\sim2351keV$ energy channels respectively,
 in which the value of Signal over
 average background is equal or larger than 2. The black star denotes the position of the
 epicenter of Chile Earthquake. (c)The disturbance of the VLF electric spectrum in ionosphere over
the epicenter detected by the DEMETER satellite from the same
orbit 30094. \label{f30094}
 }
}
 \end{figure*}

In Figure.~\ref{f30109}, plot (a) displays the distribution of
high energy charged particle average counting rates in 2010
(~\cite{Zhangzhenxia2012}), from which we can see the peak of the
average counting rates very clearly shown on 16th, February in the
lower energy region of 90$\sim$600keV, named PB1, coming from
upward orbits data of 30109.  The distribution of the flux
enhancement of the PBs with two times flux over the averaged value
of background is shown in plot (b). The selected criteria is:
$signal/BG\ge2$, in which signal denotes the real flux in every
pixels and BG denotes the averaged value of flux of all revisit
orbits in the first three months of year 2010 over the same focus
region. The disturbance of the electric field in ionosphere over
the same region in the same orbit number 30109 detected by the
DEMETER satellite is shown in plot (c).  In Figure.~\ref{f30094},
we present the the distribution of high energy charged particle
average counting rates from the orbit number 30094, the position
distribution of the flux enhancement, and the corresponding
disturbance of the electric field in ionosphere over the same
region with the same orbit number 30094. The peak of the average
counting rates appear in energy channel of 600$\sim$1000keV and
1000$\sim$2350keV from upward orbits data, which is named PB2.

 Both two PBs for different energy channels
exceed about 4 to 6 times over the average value of the counting
rates. The obvious particle counting rates enhancement also appear
in the northern hemisphere mirror points conjugate of epicenter,
induced by the bounce motion of particles between north and south
pole along the geomagnetic field line.

Corresponding to 30109 orbit on 15th February, the frequency range
of the electric field disturbance is around 14 to 20 kHz, and
corresponding to 30094 orbit on 16th February, the frequency range
of the disturbance is less than 100Hz.

 Table.~\ref{timeoccur} 
 show the observation time contrast for
 the PBs and the VLF electric enhancements.
From this table ,we can easily find there exits strong temporal
correlation between the energetic particle bursts and the VLF
electric field disturbance, so they are likely to origin from the
same VLF EM source, then the wave-particle interaction takes place
in ionosphere at the altitude of DEMETER satellite.

\begin{table*}
\tabcolsep=15pt  
\small
\renewcommand\arraystretch{1.2}  
\begin{minipage}{15.5cm}{
\small{\bf Table 1.} The parameters contrast of high energy
particle bursts and the corresponding VLF electric
 distribution(VLF). Here the pitch angle
 denotes the equatorial pitch angle of two PBs. PB1 denotes
 particle burst from 30109 orbit, PB2 denotes that from 30094
 orbit.
}
\end{minipage}
\vglue5pt
\begin{tabular}{c|c|c|c|c}
 \hline
 \hline
　event&energy/frequency&time(UT)&time(UT)& Pitch Angle($^{\circ}$)\\
\hline
    PB1&90-200keV&&03:04:30-03:15:30&21.1-29.8\\
  \hline
    PB2&150-1000keV&02:40:22-02:49:42& &21.4-30.1\\
    \hline
    PB2&1-2.35MeV&02:40:46-02:48:02& &21.4-30.1\\
    \hline
    VLF1 &$<100Hz$&02:43:42-02:50:00& &\\
    \hline
    VLF2 &14-20kHz&&03:12:00-03:14:00  &\\
 \hline
 \hline
  \end{tabular}\label{timeoccur}
\end{table*}

\newpage

\section{Spectrum of PBs}
\begin{figure*}
\center{
 \noindent\includegraphics[width=36pc]{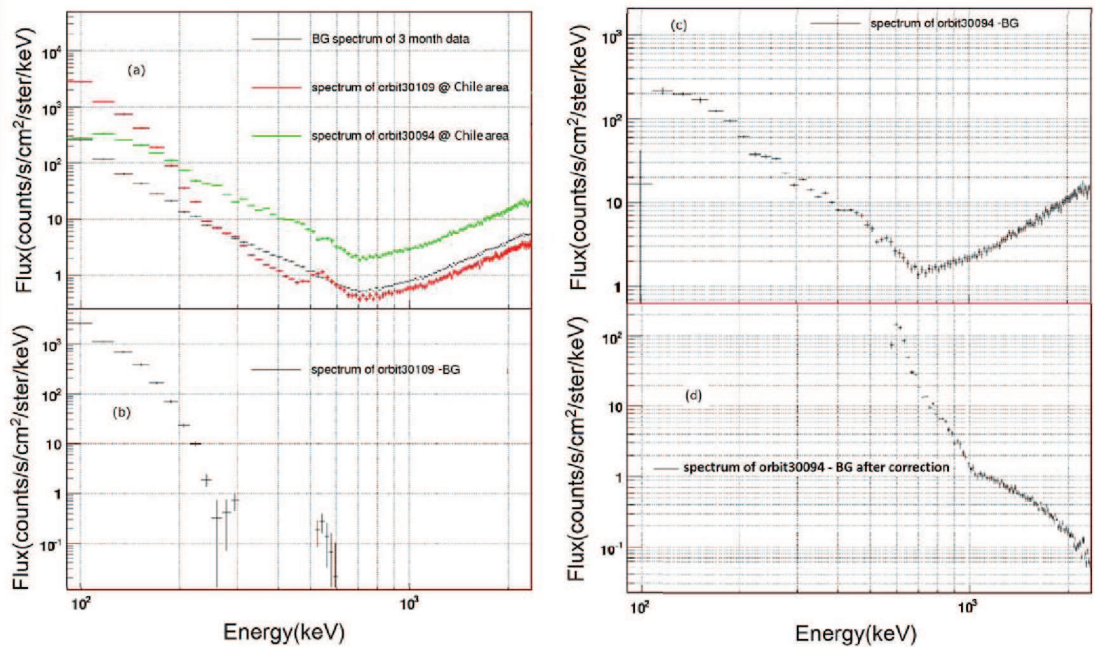}
 \caption{Energy spectrum of PBs and their background. (a)The black profile
graph with error bar denotes the average energy spectrum for the
first three months of 2010 except that of 15th and 16th February,
and the red one for that of 16th February(30109 orbit), the green
on for 15th February(30094 orbit). (b)Energy spectrum of PB1 from
30109 subtracting that of background. (c)Energy spectrum of PB2
from 30094 subtracting that of background. (d)Energy spectrum of
PB2 from 30094 subtracting that of background, after performing
the corrections of efficiency and incident-deposition energy
relation of proton. \label{energyspc}
 }
 }
 \end{figure*}

In order to study the energy spectrum characteristic of the PBs
described above, we select the data sample in the first three
months of year 2010 over the epicenter region of Chile EQ
eliminating the data of 30109 and 30094 orbits, then take it as
the background data and compare the energy spectrum of the PBs to
the background, shown in Figure.~\ref{energyspc}. 

From the energy spectrum profile, it can be confirmed that the
energy of particle counting rates enhancement from 30109 orbit
focus on $0.09\sim0.2$MeV, and that from 30094 orbit focus on
$0.15\sim2.35$MeV. While the corresponding VLF electric spectrum
disturbances coming from the same orbits to the PBs have the EM
frequency of $14\sim20$kHz and less than 100Hz, respectively. The
details are concluded in the Table.~\ref{timeoccur}.

There are very big difference for the frequency of VLF wave in the
wave-particle interaction between the two couplings, one
distributes in 14-20kHz, another in less than 100Hz. Under the
same condition of the L value, the equatorial pitch angle, the
longitude and latitude and so on, there must be a certain
mechanism driving the coupling process, for instance, the two PBs
may origin from different particle species: electron and proton.
IDP has no ability to identity the particle species, so we have to
verify this assumption by theory calculation.

Since the detected electron efficiency is very low in the energy
range larger than 1MeV, we assume that the particles of PB2 are
likely to coming from the protons. This supposition is proved to
be correct in the theory calculation analysis in the next section.
 The science data of
level-1 was reconstructed in the condition of thinking all the
particles as electrons, so we need recalculate the incident energy
spectrum for the PB2 particles by taking them as protons. In IDP
the optic has an aluminum foil with a thickness of 6$\mu$m to stop
protons with energies lower than 500keV(~\cite{Sauvaud2006}). The
detected energy spectrum can be derived by counting for the
electron efficiency profile. The detected energy spectrum is the
same as the incident energy spectrum of proton, since the
efficiency of proton penetrating through the aluminum foil is
about $100\%$. So the corrected incident energy range for the
particles in PB2 is derived to be $0.65\sim2.85$ MeV , shown in
Figure.~\ref{energyspc}(d).

\begin{figure}
\center{
 \noindent\includegraphics[width=20pc]{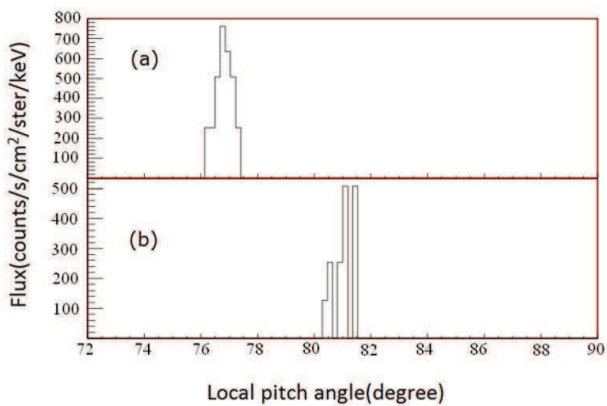}
 \caption{The local pitch angle
 distribution of the PBs. (a) for PB1 from 30109 orbit, (b)for PB2 from 30094 orbit. \label{pitchangle}
 }
}
 \end{figure}

IDP has no angular identification ability to the incident
particles, but fortunately they recorded the pitch angle by
calculating the direction between the axis of IDP aperture and the
local magnetic field magnetic field. The pitch angle for the PBs
observed here are presented in Figure.~\ref{pitchangle}. 
The particle pitch angles are distributed within
$76^{\circ}\sim78^{\circ}$ and  $80^{\circ}\sim82^{\circ}$ for the
two
PBs respectively ( see Table.~\ref{timeoccur} 
 for details). At the latitude of $32.8^{\circ}\sim39.8^{\circ}$ corresponding to the
selected region above Chile earthquake, the equatorial pitch
angles are derived to be: $21.1^{\circ}\sim29.8^{\circ}$ and
$21.4^{\circ}\sim30.1^{\circ}$.

\section{Verification of Pitch Angle Diffusion Equation in Wave-Particle Interaction Theory}

Despite the magnetospheric physics has been studied for more than
40 years, the dynamics of the Earth's particle radiation belts are
not well understood and has been attracting people's interest to
research it. Among many papers discussing the processes governing
the transportation, acceleration, and loss of radiation belt
electrons, what engage people's attention is the paper
(~\cite{Gendrin2001,Horne2002}) which assesses the influence of
wave-particle interactions on radiation belt electron dynamics. In
the early work(~\cite{Kernel1966}) on pitch angle scattering of
radiation belt particles, the gyroresonant wave-particle
interaction was thought of playing a crucial role in
magnetoshperic physics. The work of Summers(~\cite{Summers2005})
develops a special form of the the quasi-linear diffusion
coefficients corresponding to(R-mode or L-mode) electromagnetic
waves with a Gaussian spectral density propagating in a hydrogen
plasma. They derived a collisionless Vlasov equation for the
particle distribution function to the second order  in
perturbation and obtained the diffusion equation. In the inner
radiation belts, the particle velocities generally are much larger
than typical phase velocities of waves, so pitch angle diffusion
plays a dominant role in the wave-particle interaction.

Originating from the Fokker-Planck equation(~\cite{Melrose1980}),
the quasi-linear diffusion equation with only pitch-angle
diffusion coefficients for cyclotron resonant interaction with
field-aligned electromagnetic waves of arbitrary spectral density
can be written as:
\begin{equation}
\frac{\partial f}{\partial
t}=\frac{1}{\sin\alpha}\frac{\partial}{\partial
\alpha}(D_{\alpha\alpha}\sin\alpha\frac{\partial
f}{\partial\alpha}),
\end{equation}
where, for one particle of certain species $\sigma$, charge $q$
and rest mass $m_{\sigma}$, $f(\alpha,E,L)$ is the density
function in the gyrophase-averaged phase space, which depends on
the equatorial pitch angle $\alpha$, the kinetic energy E, and the
McIlwain parameter $L$.

Assuming the wave frequency spectrum density obeys Gaussian
distribution, the local pitch angle diffusion coefficient of
wave-particle interaction is deduced as following:
\begin{eqnarray}
D_{\alpha\alpha} & =&
\frac{\pi}{2}\frac{1}{\nu}\frac{\Omega^2_{\sigma}}{|\Omega_e|}\frac{1}{(E+1)^2}\sum_s\sum_j\frac{R(1-\frac{x\cos\alpha}{y\beta})^2|F(x,y)|}{\delta
x|\beta \cos\alpha-F(x,y)|} \nonumber\\
&& \cdot  e^{-(\frac{x-x_m}{\delta x})^2},
\end{eqnarray}
where E is the dimensionless particle kinetic energy given by
$E=E_k/(m_{\sigma}c^2)=\gamma-1$;
$\beta=\nu/c=[E(E+2)]^{1/2}/(E+1)$;B is the Earth's magnetic field
and $R=|\delta B_s|^2/B^2_{0}$ is the ratio of the energy density
of the turbulent magnetic field to that of the background
field.$x_m=\omega_m/|\Omega_e|$, $\delta x=\delta
\omega/|\Omega_e|$, s=1 for R-mode wave and s=-1 for L-mode wave.
$j=1,2,\cdots,N $ is the root number of satisfying the resonance
condition,
\begin{equation}
\omega_j-\nu \cos\alpha
k_j=-s\frac{q}{|q|}\frac{|\Omega_{\sigma}|}{\gamma},
\end{equation}
$F(x,y)=dx/dy$ ($x=\omega/|\Omega_e|, y=ck_i/|\Omega_e|$) is
determined from the dispersion equation of electron or proton as
following:
\begin{equation}
(\frac{ck}{\omega})^2=1-\frac{(1+\epsilon)/\alpha^{\ast}}{(\omega/|\Omega_e|-s)(\omega/|\Omega_e|+s\epsilon)},
\end{equation}
where
\begin{equation}
\alpha^{\ast}=\Omega^2_e/\omega^2_{pe}
\end{equation}
is an important cold-plasma parameter;$\epsilon$ is the rest mass
ratio of electron and proton; $|\Omega_e|=e|B_0|/(m_ec)$ denotes
the electron gyrofrequency and $\omega_{pe}=(4\pi
N_0e^2/m_e)^{1/2}$ is the plasma frequency, where $N_0$ denotes
the particle number density in ionosphere at the altitude of
Satellite observation.

From the VLF electric spectrum distribution and energy spectrum of
PBs, 
 the typical
wave-particle interaction takes place in the ionosphere.
Furthermore, the coupling energy of particle and the frequency of
VLF wave take on specific correlation, as listed in Table.~\ref{timeoccur}.

Assuming the wave-particle interaction occurs only within the
equatorial plane, we can calculate the equatorial pitch angle
diffusion coefficient for the electron coupling with R-mode
electromagnetic wave and the proton coupling with L-mode
electromagnetic wave. The parameters are chosen as following: the
wave amplitude $\delta b= 10$ pT, the equatorial magnetic field
determined by a dipole model $B=3.11\times 10^{-5}/L^3 $ T. The
equatorial plasma density $N$ is equal to $N_0 \times (2/L)^4$
cm$^{-3}$ according to the study(~\cite{Angerami1964,Inan1984}).
Due to the location of PBs in the South Atlantic Anomaly where the
average particle counting ratio is much larger than other region
in ionosphere, we select particle density $N_0=18000 $cm$^{-3}$
which is at least reasonable in the magnitude comparing to the
satellite detection result. The other parameters used in the
theory calculation can be found in every plot.

\begin{figure}
 \noindent\includegraphics[width=18pc]{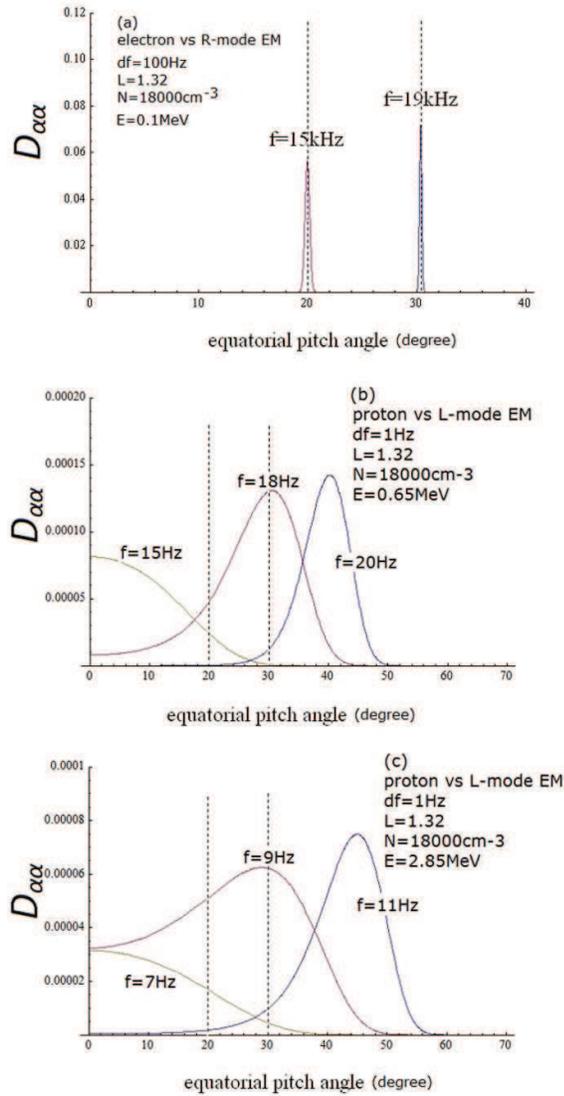}
 \caption{The dependence of pitch angle diffusion coefficient distribution on equatorial pitch angle.
 The parameters used here are: L value 1.32 and particle density $18000 cm^{-3}$. (a) for the electron coupling with R-mode EM wave with the frequency bandwidth 100Hz.
   (b) for the 0.65MeV proton coupling with L-mode EM wave.
  (c) for the 2.85MeV proton coupling with L-mode EM wave.
   The vertical dashed lines denote the satellite observation range in pitch angle.
   \label{Dalpha}
 }
 \end{figure}

Eventually, we obtained the nice agreement between the DEMETER
observation and the qusi-linear diffusion equation calculation, as
long as the PBs of 30109 is explained to origin dominantly from
electron and the PBs of 30094 from proton, that is, the electron
couples with R-mode EM wave and the proton with L-mode EM wave in
the wave-particle interaction mechanism. Figure.~\ref{Dalpha} show
the profile of the equatorial pitch angle diffusion coefficient
$D_{\alpha\alpha}$ depending on the pitch angle distribution.

\newpage

In Figure.~\ref{Dalpha}, 
 for the coupling of electron and
R-mode EM wave, we choose the bandwidth of wave spectrum with
$df=100Hz$, and for that of proton and L-mode EM wave, the
frequency bandwidth is df=1Hz. The L value with $L=1.32$ in the
center of the observation region is selected. The range between
two dashed lines in the figure indicates the observation limit of
DEMETER satellite. We can find the pitch angle of wave-particle
interaction with the electron energy 0.1MeV and VLF EM wave
frequency from $15\sim19$kHz distributes in around
$20^\circ\sim30^\circ$ and just enter the sight of IDP. As for the
proton vs L-mode EM wave, considering the parameters range of PBs
and the VLF EM wave observed, we select several cases with
different interaction parameters and calculate the pitch angle
diffusion coefficient. The protons with energy 0.65MeV interact
with VLF wave with frequency of 15Hz,18Hz,20Hz. And the protons
with energy 2.85MeV interact with wave of 7Hz,9Hz,11Hz. Their
pitch angle diffusion can completely or partially fall into the
observed range of equatorial pitch angle around
$20^\circ\sim30^\circ$ between the two dashed line in the plots.

\begin{figure*}
\center{
 \noindent\includegraphics[width=36pc]{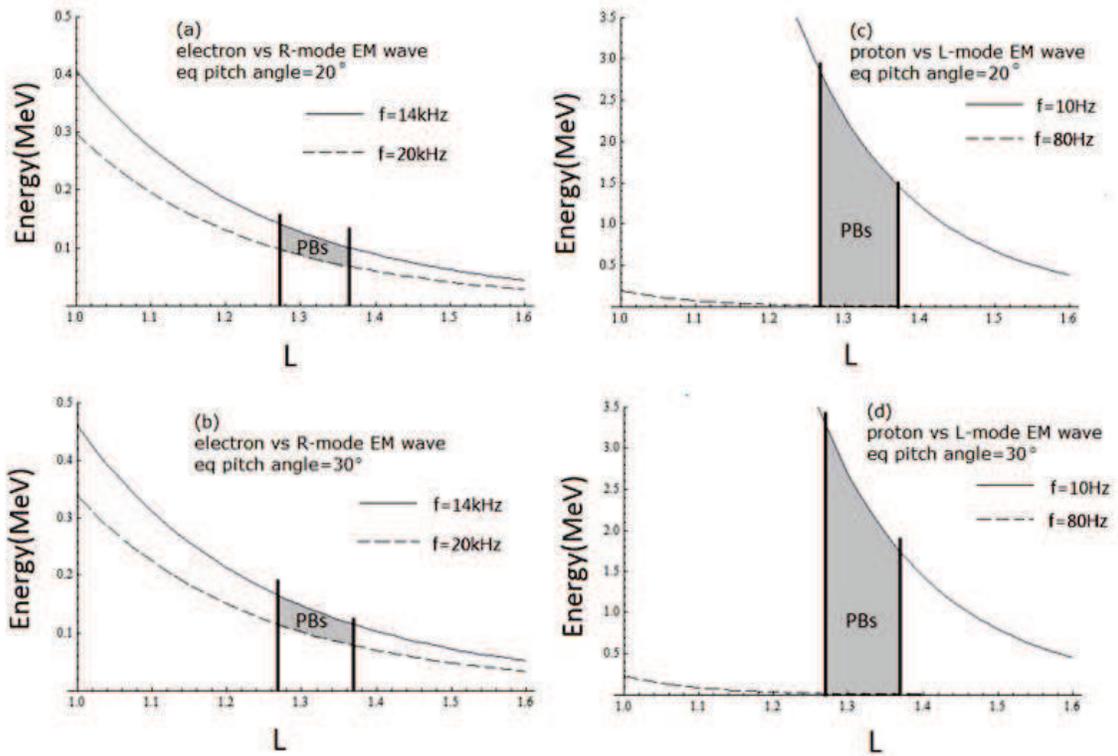}
 \caption{The particle energy calculated by quasi-linear diffusion coefficient equation, shown in the shadow region.
 Parameter L ranges from 1.27 to 1.37, just in accordance with the range of PBs observed. The solid and dashed curves represent
 the different wave frequency used in the calculation.
 \label{energyL}
 }
}
 \end{figure*}

Under the condition of a given equatorial pitch angle $\alpha$, EM
frequency f and other parameters, the particle energy of the
quasi-linear diffusion coefficient theory relies on the L value.
The reports showed a decrease in energy with increasing L from
cyclotron resonance(~\cite{Koons1981,Chang and Inan1983}). In
Figure.~\ref{energyL},
 we present the
coupled energy of electron and proton depending on L value by
theoretical calculation, (a)and (b) for electron coupling with
R-mode EM wave, (c) and (d) for proton coupling with L-mode EM
wave. The equatorial pitch angles in (a) and (c) used in the
theoretical calculation are selected to be: $20^\circ$ and
$30^\circ$ in (b)and (d), which is just consistent with the range
of observed PBs listed in the Table.~\ref{timeoccur}.
 We find
that, within the range of L $1.27\sim1.37$, the energy of particle
calculated corresponding to the shadow region in each plot is
almost consistent with the energy of the PBs observed by DEMEMTER,
that is ,$0.1\sim0.2$MeV for electron and $0\sim3.4$MeV for
proton.

\begin{figure*}
\center{
 \noindent\includegraphics[width=36pc]{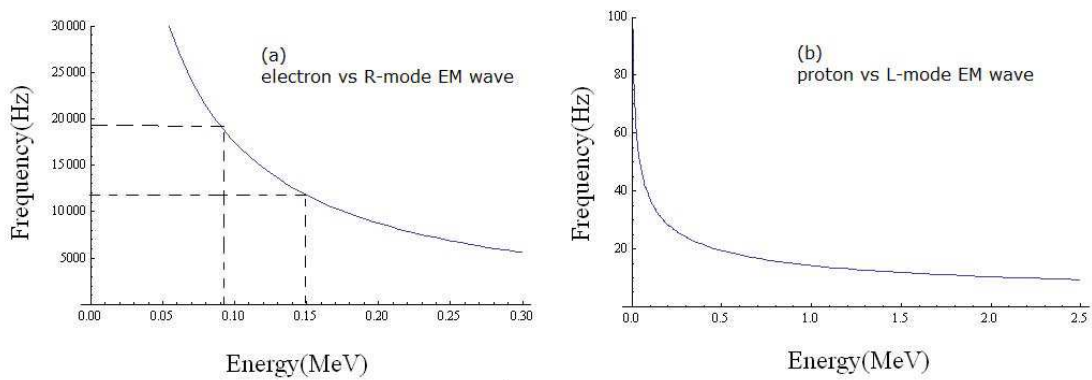}
 \caption{The coupling relation of frequency of EM wave and the energy of particles
  calculated by the theory of quasi-linear diffusion coefficients,
  (a) for the electron coupling with R-mode EM wave, (b) for the
  proton coupling with L-mode EM wave.
 \label{energyfrequency}
 }
}
 \end{figure*}

The relation between the particle energy and the VLF EM wave in
the wave-particle interaction are shown in the
Figure.~\ref{energyfrequency},
 derived by the resonance condition (3)and
the dispersion equation (4).

 For electron coupling with R-mode EM wave, the particles with
 energy of 0.09MeV-0.15MeV interact with VLF EM wave with the
 frequency of $12\sim20$kHz. For proton vs L-mode wave, particles with energy of $0.65\sim2.85$ MeV  interact with VLF wave
 with the frequency of less than 100Hz. This calculation results
 agree mostly well with the observation range of PBs and VLF electric disturbance, seeing the detail in Table.~\ref{timeoccur}.

\section{Discussion and Summary}
For the study of the high energy particle counting rates
enhancement and VLF electric disturbance observed simultaneously
in ionosphere,
we analyze the energy spectrum of PBs and show the pitch angle
distribution of them and point out that the typical wave-particle
interaction take place, which induces the energetic charged
particle entering the drift loss cone, causing the particle
counting rates enhancement locally in ionosphere. The energy range
for the two PBs distributes around $0.09\sim0.2$MeV and
$0.65\sim2.85$ MeV , but the frequency of VLF wave coupling with
particles has much big different values $14\sim20$kHz and less
than 100Hz.

If we regard the particles in PBs as electron and proton
respectively, the agreement with observation is obtained by
calculation of pitch angle quasi-linear diffusion coefficient
equation. We verify the supposition by checking three kinds of
correlation formula: pitch angle diffusion coefficient
$D_{\alpha\alpha}$ depending on pitch angle $\alpha$, the coupled
particle energy depending on McIlwain parameter L and the relation
between coupled particle energy and VLF wave frequency. So it
confirms that the PBs and VLF electric spectrum disturbance are
caused by a same VLF EM wave source, and proves that we find one
typical wave-particle interaction based on DEMETER satellite.

For the first time, under the inability to identify particle
species of IDP on board DEMETER, by wave-particle interaction
calculation, we point out the particle bursts origin accurately
from electron or proton dominant source.
 In addition, the PBS and electric spectrum disturbance appearing
 simultaneously over the Chile EQ epicenter about 10 days before the
 earthquake,
  if there is really certain link
 between the PBs and VLF electric disturbance
and seismic activity as argued by the
paper(~\cite{Zhangzhenxia2012}), then this is also the first time
that we obtain the evidence of precursor information of seismic
activity, that is, the electron and proton counting rate
enhancement in ionosphere simultaneously observed by satellite.

The consistent results with observation also prove the correctness
and rationality of the theory of quasi-linear diffusion
coefficient equation applying in the particle precipitation and
wave-particle interaction analysis in the low orbital satellite
observation in ionosphere, in turn. It is worth saying that the
errors, associated with the negligence of higher-order resonances,
are smaller than inaccuracies associated with uncertainties in the
input values for the plasma density and latitudinal distribution
of the waves. Although the reasonable assumption have been made,
there is still some limitation and a long road to describe
precisely the observation for the theory. Such as, the accuracy of
the plasma density playing an important role in theory always
depend on the more accurate measurement of satellite detection to
ionosphere. Many researchers studied the outer radiation dynamics
of 2D momentum-pitch-angle diffusion equation of energetic
electrons interaction with whistler wave, and the translation
process of whistler wave during a little strong geomagnetic
activities(~\cite{Zheng2008,xiaofl2011,suzp2009}), and there are
still other related works(~\cite{jiang2011,zhang2014}) focusing on
the property of wave propagation and wave-particle interaction. So
we will continue further study based on those works and attempt to
perform some quantitative calculation.

Laboratory measurements indicate that rock samples under stress
can radiate EM energy in a wideband spectrum, and that light can
been observed(~\cite{Cress1987}). Estimations of the size of the
EQ preparation zone are thought to be ranging from 150 km for
magnitude 5 EQs to more than 2,500 km for magnitude
8(~\cite{Dobrovolsky1979}). So if it is true that the seismic
activity radiate EM waves and the waves can propagate into the
ionosphere and diffuse the energetic particles by wave-particle
interaction, then we can monitor the particles change to
investigate the seismic activity. So we need to explore the
theoretical model to provide the support for this hypothesis.

In this paper, we only perform qualitative analysis for the EM
waves coupling with energetic particles in ionosphere. In order to
systematically quantitatively research seismic electromagnetic
radiation in lithosphere-atmosphere-ionosphere and the caused
ionospheric disturbance, in the future we will still apply the
proper propagation model of EM wave, such as ray tracing method,
for calculating the power reduction of waves and the wave normal
angles and fields of whistler mode waves in the
ionosphere(~\cite{Starks2009,Tao2010}). In the process, we will
consider using the parameters more close to the real condition,
including more real geomagnetic field model rather than the dipole
magnetic field, non-uniform magnetic flux density and more real
plasma layer density. In term of wave-particle interaction, we
will consider the effect of radial diffusion of particle and apply
the non-linear cyclotron resonant interaction of energetic
electrons and coherent VLF waves, to quantitatively calculate the
precipitation or space distribution change of energetic
particles(~\cite{Inan1978}).

Although the position we select the data to study locates in the
South Atlantic Anomaly(SAA), the saturation of IDP detector does
not take place.
 In previous paper(~\cite{Zhangzhenxia2012}, we show the distribution of high
energy charged particle counting rates in the epicenter region for
three different energy regions, from left to right plots are the
distribution for the year 2007, 2008 and 2009. From these figures,
we can find the maximum value of averaged counting rates in the
first plot of $90.7\sim600$keV @2007 is close to $300\times 10^3$
, while the maximum value of the particle burst is only less than
$120\times 10^3$ . So there is not saturation of IDP detector
above SAA.

The energy spectra in paper(~\cite{Zhangzhenxia2012} 
 with the flux increase
above 700 keV, does not indicate a pile-up
  effect of the electronics. The shape of spectra with increase above 700 keV is due to the efficiency
  correction(~\cite{Sauvaud2006}).
    In addition, this case of energy spectrum shape is also the same when the flux is very low.
    The data we download from DEMETER website is the science data
    of level-1 after performing efficiency correction.
     So basically, there is no pile-up for the data we used. The data we used in this work is reliable.

\begin{acknowledgments}
The authors would like to express their sincere thanks for the
provision of data download of DEMETER Project. This work was
supported by Spark Plan for Earthquake Science and Technology of
China Earthquake Administration(Grant No XH12066) and National
Natural Science Foundation of China(Grant No 11103023) and
Director Foundation of National Earthquake Infrastructure
Service(2014) .
\end{acknowledgments}

\end{article}

\end{document}